\documentclass[12pt]{iopart}
\usepackage{graphicx}
\usepackage{iopams}  
\usepackage{amssymb}
\usepackage{color}
\usepackage{bm}      % Lettre grasses
\usepackage{booktabs}
\usepackage{array}
\usepackage{lipsum}

\newcommand{\bea}{\begin{eqnarray}}
\newcommand{\eea}{\end{eqnarray}}
\newcommand{\beq}{\begin{equation}}
\newcommand{\eeq}{\end{equation}}
\newcommand{\bit}{\begin{itemize}}
\newcommand{\eit}{\end{itemize}}

\newcommand{\ra}{\rightarrow}
\newcommand{\s}{\sigma}
\newcommand{\ot}{ \omega}
\newcommand{\D}{\mathrm{d}}

\renewcommand{\P}{\mathcal{P}}

\renewcommand{\arraystretch}{1}

\newcolumntype{C}{ >{\centering\arraybackslash} m{13.5cm} }

\begin{document}

\title{Work statistics in stochastically driven systems}

\author{Gatien Verley$^1$, Christian Van den Broeck$^2$, Massimiliano Esposito$^1$}
\address{$^1$ Complex Systems and Statistical Mechanics, University of Luxembourg, L-1511 Luxembourg, G.D. Luxembourg}
\address{ $^2$ Hasselt University - B-3590 Diepenbeek, Belgium }


\date{\today}

\begin{abstract}
We identify the conditions under which a stochastic driving inducing energy changes on a system coupled to a thermal bath can be treated as a work source. When these conditions are met, the work statistics satisfies the Crooks fluctuation theorem traditionally derived for deterministic drivings. We illustrate this fact by calculating and comparing the work statistics for a two-level system driven respectively by a stochastic and a deterministic piecewise constant protocol.
\end{abstract}

\maketitle

%%%%%%%%%%%%%%%%%%%%%%%%%%%%%%%%%%%%%%%%%%%%%%%%%%%%%%%%%%%%%%%%%%%%%%%%%%%%%%%%%%%%%%%%%%%%%%%%%%%%%%%%%%
\section{Introduction}

Stochastic thermodynamics allows the identification of thermodynamic quantities at the level of single stochastic trajectories. This theory is particularly relevant for small systems in which fluctuations are comparable to average values. Remarkable experimental advances, allowing to measure and manipulate systems at the (sub) micro-size level, have validated these theoretical developments. Among these achievements are the out of equilibrium versions of the fluctuation-dissipation theorem \cite{Martin2001_vol98, Speck2006_vol74, Blickle2007_vol98, Baiesi2009_vol103, Gomez-Solano2009_vol103, Prost2009_vol103, Chetrite2009_vol137, Seifert2010_vol89, Gomez-Solano2011_vol297, Verley2011_vol}, the connection between thermodynamics and information theory \cite{Kawai2007_vol98, Berut2012_vol483, Sagawa2010_vol104, Toyabe2010_vol6, VandenBroeck2010_vol6, Esposito2012_vol99, Esposito2011_vol95, Diana2013_vol87, Barato2013_vol}, and -maybe most importantly-  the formulation and verification of the so-called fluctuation theorem (and its variants) \cite{Evans1994_vol50, Gallavotti1995_vol74, Jarzynski1997_vol78, Kurchan1998_vol31, Lebowitz1999_vol95, Crooks2000_vol61, Esposito2010_vol104, Bochkov2013_vol56}.  According to this theorem, a (stochastic) positive entropy production is exponentially more likely to be observed than the corresponding negative entropy production. On average this implies a positive entropy production in agreement with  the second law. Depending on the specific setup, entropy production can be related to physical observables \cite{BulnesCuetara2014_vol}. For instance, when considering transitions between equilibrium states for a system in contact with a single heat bath and driven by a deterministic time-dependent force, the entropy production can be expressed in terms of the work $W$ minus the free energy difference $\Delta F$ between the initial and final equilibrium state. The fluctuation theorem for the work statistics in such setups has been used to evaluate free energy differences between molecular equilibrium states \cite{Collin2005_vol437, Alemany2012_vol8}.

Stochastic driving is an issue of both conceptual as well as practical interest. Conceptually, one may wonder whether the energy transferred to the system can be treated as work. From a practical perspective a deterministic driving is never perfect and will always be accompanied by uncontrollable small random fluctuations. It is therefore important to understand how these fluctuations may affect the work statistics and the fluctuation theorem \cite{Gomez-Solano2010_vol89, Cadot2008_vol66, Falcon2009_vol79, Falcon2008_vol100}. The entropy production in the joint space of the driven system as well as of the stochastic driving will always satisfy a fluctuation theorem. This was recently shown in Ref. \cite{Michel2012_vol85} by considering Ornstein-Uhlenbeck processes. In the present paper, we want to use stochastic thermodynamics to determine the conditions under which the energy injected in the system by the stochastic driving fully captures the joint system dissipation and can thus be treated as work. We show that these conditions are rather restrictive since the stochastic driving must evolve reversibly. We study two models where this condition is satisfied: a two-level Markov process coupled to another independent two-level Markov process, and a one-dimensional Ornstein-Uhlenbeck process coupled to another independent one-dimensional Ornstein-Uhlenbeck process. We also use the former model to compare the large deviation properties of its work statistics with that of the same system but driven by a deterministic piecewise constant driving. We find that while the large deviation function has a compact support in the deterministic case, it does not in the stochastic case. Furthermore, we show that both drivings give rise to a Crooks fluctuation theorem. 

The outline of the paper is as follows. In section \ref{sec1}, we identify the general conditions under which a stochastic driving behaves as a work source. In this case we show that the work statistics satisfies a Crooks fluctuation theorem \cite{Crooks2000_vol61}. In section \ref{sec:2level}, we compute and compare the large deviation properties of the work statistics of a two-level system driven by a piecewise constant deterministic driving \cite{Verley2013_vol88} with that of the same system subjected to a stochastic work source. Conclusions are drawn in section \ref{sec:Concl}. Finally, using the results of Ref. \cite{Pal2013_vol87}, we show in \ref{sec:BrownPart} that a stochastically driven overdamped colloidal particle in a harmonic trap does satisfy a Crooks fluctuation theorem when the work is properly identified. 

%%%%%%%%%%%%%%%%%%%%%%%%%%%%%%%%%%%%%%%%%%%%%%%%%%%%%%%%%%%%%%%%%%%%%%%%%%%%%%%%%%%%%%%%%%%%%%%%%%%%%%%%%%
\section{Stochastic driving as a work source} \label{sec1}

We consider a stationary Markovian dynamics on a bipartite joint system made of a system with states $\s$ and an independent energy source with states $h$. The bipartite property means that transitions involving a simultaneous change in $\s$ and $h$ are not allowed. The rates $\omega_{\s',\s}(h)$ describing system jumps from $\s$ to $\s'$ satisfy local detailed balance \cite{Seifert2012_vol75, Esposito2012_vol85}
\beq
 \ln \frac{\omega_{\s',\s}(h)}{\omega_{\s,\s'}(h)} = - \beta [E(\s',h) - E(\s,h)], \label{eq:DBsys}
\eeq
while the rates describing the energy source $\omega_{h,h'}$ do not depend on $\s$ and do not necessarily satisfy local detailed balance. We introduced the inverse temperature $\beta=1/T$ ($k_b=1$). Such dynamics can be viewed as a limit of a global dynamics satisfying local detailed balance when the energy scale involved during the source transitions is very large compared to the system one (see \ref{LimEnSource}). The energy changes due to transitions between $\s$ states at fixed $h$ constitute the heat exchanged with the bath while the energy changes due to transitions between $h$ states at fixed $\s$ constitute system energy changes due to the energy source. 

We now turn to energy exchanges described at the level of single trajectories. To set the notation, we denote the joint system, the system, and the energy source trajectories during a time interval $[0,t]$ respectively by $[\s,h]$, $[\s]$ and $[h]$. Similarly, the time reversed trajectories are $[\bar \s, \bar h]$, $[ \bar \s]$ and $[\bar h]$. The trajectory probabilities $\P[\s,h]$ can naturally be expressed as $\P[\s,h]=\P[\s|h] \P[h]$. Thanks to the independence of the work source with respect to the system, the probability $\P[h]$ of a trajectory $[h]$ is that of the Markovian dynamics of the source solely determined by the rates $\omega_{h,h'}$. The probability $\P[\s|h]$ (resp. $\P_{\s_0}[\s|h]$) is the probability of a system trajectory $[\s]$ (resp. $[\s]$ starting in the initial state $\s_0$) subjected to a given trajectory $[h]$ of the energy source. Using the local detailed balance (\ref{eq:DBsys}), the heat entering the system from the bath for a given trajectory $[h]$ can be expressed as 
\beq
Q [\s | h ] = - \beta^{-1} \ln \frac{\P_{\s_0} [\s|h] }{ \P_{\bar \s_0}[\bar \s | \bar h ]} .
\eeq
Since $\P [\s | h ] = p(\s_0|h_0) \P_{\s_0} [\s | h ]$ where $p(\s_0|h_0) $ is the probability to be in the initial state $\s_0$ for a given initial $h_0$, the entropy production of the system is
\beq
 \Delta_{\rm i} S[\s|h] = - \beta Q [\s|h] + \Delta S[\s|h] = \ln \frac{\P [\s|h] }{ \P[\bar \s | \bar h]} ,\label{eq:ddEntProd}
\eeq
where $\Delta S[\s|h] = \ln p(\s_0|h_0) / p(\s_t|h_t)$ is the change in the Shannon entropy of the system after a time $t$. Using energy conservation along a system trajectory, the energy provided by the energy source to the system reads $W [\s|h] = \Delta E[\s,h]-Q [\s|h]$, where $\Delta E[\s,h] = E(\s_t,h_t)-E(\s_0,h_0)$. Therefore, the entropy production becomes
\beq
 \Delta_{\rm i} S[\s|h] = \beta ( W[\s|h] - \Delta F[\s|h] ) ,\label{eq:ddEntProdBis}
\eeq
where $\Delta F[\s|h]=\Delta E[\s,h] - \beta^{-1} \Delta S[\s |h]$ is the change in the nonequilibrium free energy of the system. The entropy production (\ref{eq:ddEntProd}) and (\ref{eq:ddEntProdBis}) is identical to that of a system subjected to a deterministic driving $[h]$ made of sudden jumps \cite{Seifert2012_vol75, Esposito2010_vol82, VandenBroeck2010_vol82, Broeck2013_vol}. However, since the energy source is stochastic and produces a statistical ensemble of drivings $[h]$, the entropy production of the energy source $\Delta_{\rm i} S_{sd}=\ln \P[h]/ \P[\bar h]$ gives rise to an additional contribution to the joint system entropy production
\beq
 \Delta_{\rm i} S[\s,h] = \Delta_{\rm i} S[\s|h] + \ln \frac{\P[h]}{ \P[\bar h]}. \label{eq:TotEntProd}
\eeq
Ensemble averaging (\ref{eq:TotEntProd}), we get
\beq
\langle \Delta_{\rm i} S[\s,h] \rangle = \sum_{[h]} \P[h] \sum_{[\s]} \P[\s | h ] \Delta_{\rm i} S[\s|h] + \sum_{[h]} \P[h] \ln \frac{\P[h]}{\P[\bar h]} \geq 0. \label{eq:MeanTotEntProd}
\eeq
Since a work source is not supposed to give rise to any entropy production, our energy source is only a work source when this additional term vanishes and thus does not affect the system entropy balance. This happens either when the driving is deterministic or when the energy source $h$ evolves reversibly, i.e. when $\P[h]=\P[\bar h]$. In this latter case, from the system perspective, the trajectory $[h]$ of the energy source is perceived as a time-dependent stochastic driving. We note that even in presence of a dissipative energy source, the non-negative first term on the r.h.s. of (\ref{eq:MeanTotEntProd}) still provides a lower bound to the entropy production of the joint system. The non-negative second term constitutes in turn a lower bound on the dissipation of the energy source, since the trajectories $[h]$ may only provide a coarse grained description of the energy source dynamics.   

When the long time limit is considered and for systems with a finite state space, the contributions $\Delta F[\s|h]$ to the entropy production (\ref{eq:ddEntProdBis}) is not extensive in time and thus vanishes in the large deviation sense. The steady state fluctuation theorem for the entropy production therefore reads
\beq
I(w+\Sigma_{sd})-I(-w-\Sigma_{sd} )=-w-\Sigma_{sd} , \label{eq:new-WFT}
\eeq
where $w=W/t$ is the energy per unit of time provided by the energy source to the system, $\Sigma_{sd} =  \Delta_{\rm i} S_{sd}/t$ is the rate of entropy production due to the energy source, and $I(w+\Sigma_{sd}) = \lim_{t\rightarrow \infty} (-1/t)\ln P(\Delta_{\rm i} S = t w+t\Sigma_{sd})$ is the large deviation function for the total entropy production.
It is once again only when the driving is deterministic, or when the energy source $h$ evolves reversibly, that $\Sigma_{sd}=0$, that the energy source behaves as a work source, and that the Crooks fluctuation theorem is recovered.

The stochastic drivings used for the models presented below all satisfy this condition and qualify as work sources. 

%%%%%%%%%%%%%%%%%%%%%%%%%%%%%%%%%%%%%%%%%%%%%%%%%%%%%%%%%%%%%%%%%%%%%%%%%%%%%%%%%%%%%%%%%%%%%%%%%%%%%%%%%%
\section{Modulated two-level system}
\label{sec:2level}

In this section, we compare the work statistics of a two-level system driven by a stochastic (reversible) work source with that of the same system driven by a deterministic work source.

%%%%%%%%%%%%%%%%%%%%%%%%%%%%%%%%%%%%%%%%%%%%%%%%%%%%%%%%%%%%%%%%%%%%%%%%%%%%%%%%%%%%%%%%%%%%%%%%%%%%%%%%%%
\subsection{Stochastic work source}

\begin{figure}[h]
\begin{center}
\includegraphics[width= 7cm]{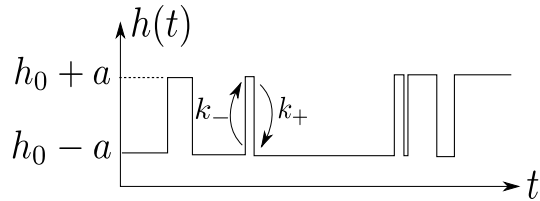}
\end{center}
\caption{Representation of the work source stochastic dynamics.}
\label{fig0bis}
\end{figure}

\begin{figure}
\begin{center}
\vspace{0.5cm}
\includegraphics[width= 7cm]{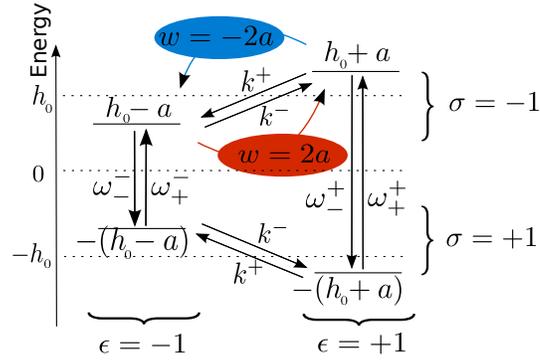}
\end{center}
\caption{Representation of the energy levels of the two-level system: on the left (resp. right) the work source state is  $\varepsilon = -1 $ (resp. $\varepsilon = +1$). The upper (resp. lower) part of the figure corresponds to $\s=-1$ (reps. $\s=1$). For a $\s=-1$, a decrease of $h$ will decrease the energy of the system resulting in a negative work contribution $w=-2a$, while an increase of $h$ will increase the energy of the system and result in a positive work $w =2a$.}
\label{fig7}
\end{figure}

We consider a two-level system $\s = \pm 1$ coupled to a heat bath at temperature $T$ and interacting with a stochastic energy source with two states $\varepsilon = \pm 1$. The Poisson rate to leave the source state $\varepsilon$ is denoted $k^\varepsilon$ and does not depend on $\s$. This immediately implies that the stationary dynamics of the energy source is reversible and can thus be considered as a work source. We denote by $h(t)=h_0+\varepsilon(t) a$ the driving produced by the work source whose state at time $t$ is given by $\varepsilon(t)$. This driving is illustrated in Fig.~\ref{fig0bis}. The energy of the joint system is $E(\s,h) = -\s h $ in unit of $k_B T$ ($\beta =1$ from now on). This corresponds to the four energy levels $E(\s,h_0+\varepsilon a) = -\s(h_0+ \varepsilon a)$ depicted in Fig.~\ref{fig7}.
For a given source state $\varepsilon$, the rate describing a transition from system state $\sigma$ to $-\sigma$ is given by $\omega_\s^\varepsilon = \omega(h_0+\varepsilon a) e^{-\s (h_0+\varepsilon a )}$. This rate satisfies the local detailed balance condition and includes as special case
\begin{itemize}
\item Arrhenius rates, $\omega(h) = \Gamma$,
\item Fermi rates, $\omega(h) = \Gamma/ (2 \cosh(h))$,
\item Bose rates, $\omega(h) = \Gamma/ (2 |\sinh(h)|) $,
\end{itemize}
where $\Gamma$ is a positive constant setting the time scale ($ \Gamma =1$ in all figures).
%%%%%%%%%%%%%%

%%%%%%%%%%%%%%%%%%%%%%%%
The dynamics of the joint system is described by a stationary four-state model. Each state is specified by the pair $(\s,\varepsilon) = \theta $. The work generating function is defined as $G_{\theta,\mu}(t) = \langle e^{\mu W(t)} \delta_{\theta,\theta(t)} \rangle$ where $\delta$ is the Kronecker delta and $\langle .. \rangle $ denotes an average over all possible values of the work $W(t) = - \int_0^t \D t' \dot h(t') \s(t')$ and of the states $\theta(t)$ at time $t$. Its evolution is ruled by 
\beq
\partial_t G_{\theta,\mu}= \sum_{\theta'} M^{\mu}_{\theta,\theta'} G_{\theta',\mu}, \label{eq:GenFuncEvoStoch}
\eeq
where
\beq
\bm{M}^\mu  = \left [\begin{array}{cccc}
-k^+-\omega_+^+  & \omega_-^+ & k^- e^{-2a \mu } &0 \\
\omega_+^+ & -k^+-\omega_-^+  & 0 & k^- e^{2a\mu} \\
k^+e^{2a\mu}& 0 & -k^- -\omega_+^-  & \omega^-_- \\
0& k^+e^{-2a\mu} & \omega^-_+ & - k^- - \omega^-_-  \\
\end{array} \right ]\label{eq:MasterEqGenStoch}.
\eeq
We note that while the joint system is ruled by an autonomous steady state dynamics in the long time limit, the two-level system continuously undergoes random energy switches from the work source and tries to relax toward the new corresponding equilibrium state. The work statistics in the long time limit is characterized by $\phi_\mu$ the largest eigenvalue of $\bm{M}^\mu$ and the asymptotic cumulant generating function for the work
\beq
G_\mu(t) = \sum_{\theta} G_{\theta,\mu} (t) \asymp \exp{\left [ t\phi_\mu\right ]}.
\eeq
Its analytical expression can be obtained using the Ferrari method to solve the following characteristic equation 
\beq
\det (\bm{M}^\mu - \phi \bm{1} ) = \phi^{4} + c_3 \phi^{3} +c_2\phi^{2}+c_1 \phi^{1} + c_0(\mu) =0, \label{eq:detPolynomeMain}
\eeq
where $\bm{1}$ is the identity matrix and 
\bea
c_3 &=& 2(k^+ + k^-) + \ot^+ +\ot^- > 0, \label{eq:cMain}  \\
c_2 &=& \Omega[\cosh(2h_0) + \cosh(2a) ]  +(k^+ + k^-)^2 \nonumber \\ && + \ot^+(2k^- + k^+) + \ot^-(2k^+ + k^-) > 0 , \nonumber \\
c_1 &=& (k^- + k^+)\left\{ k^-\ot^++k^+ \ot^-  \right. \nonumber \\ &&+ \left. \Omega \left[ \cosh(2h_0) + \cosh(2a)\right]  \right\} > 0, \nonumber \\
c_0(\mu)& =& -k^+k^-\Omega \left[ \cosh(2a(1+2\mu))- \cosh(2a) \right] \le 0, \nonumber
\eea
with $\Omega = 2 \omega(h_0+a) \omega(h_0-a) > 0 $ and $\ot^\pm = \omega_+^\pm + \omega_-^\pm $. Note that the last inequality in (\ref{eq:cMain}) holds for any real $\mu$. The characteristic polynomial can always be factorized into a product of two second degree polynomials with real coefficients. The cumulant generating function is then the largest solution of a second degree polynomials, with the choice between the two polynomials depending on the parameters. We do not explicitly provide the lengthy analytical solution of equation (\ref{eq:detPolynomeMain}). Simpler expressions can be obtained in the fast and slow modulation limit, or for small and high amplitudes of the field by pertubatively expanding the function $\phi$ and the coefficients $c$ to various orders. The characteristic polynomial obtained in this way decomposes into several equations for each order. The lowest order solutions are summarized in Table \ref{tab:limitcases}. 
%%%%%%%%%%%%%%%%%%%%%%
%% Tableau avec tous les cas limites CAS STOCHASTIQUE
%%%%%%%%%%%%%%%%%%%%%%
\begin{center}
\renewcommand{\arraystretch}{2.6}
\setlength{\tabcolsep}{0.4cm}
	\begin{table*}[h!]
 \begin{tabular}{c|C} 
$\Gamma \tau \ll 1$ & \raisebox{0.15cm}{$ \displaystyle \frac{\tau }{2}\sqrt{ \left [ \alpha \ot^- +(1-\alpha)\ot^+ \right ]^2 + 4 \alpha (1-\alpha) \ot^+ \ot^- \frac{\cosh 2a(2\mu+1) - \cosh 2 a }{  \cosh 2a + \cosh 2 h_0 }  } $} \\
& \raisebox{0.7cm}{$ \displaystyle
 -\frac{\tau }{2} \alpha \ot^- - \frac{\tau }{2} (1-\alpha)\ot^+$}  \\ 
$\Gamma \tau  \gg 1$ & \raisebox{0.15cm}{$ \displaystyle \frac{-1}{2 \alpha (1-\alpha)} + \frac{1}{2} \sqrt{\frac{1}{ \alpha^2 (1-\alpha)^2}+\frac{}{ } \frac{4 \left[ \cosh 2a(2\mu+1) - \cosh 2 a \right]  }{\alpha (1-\alpha) (\cosh 2a + \cosh 2 h_0 ) }}.$} \\
$ a \ll 1$ & \raisebox{0.15cm}{$\displaystyle \mu (1+\mu) \langle w \rangle = \frac{8a^2\mu(1+\mu)}{[\alpha(1-\alpha)\omega(h_0)\tau]^{-1} \cosh h_0 + 1+ \cosh 2h_0} $} \\
\raisebox{-0.6cm}{$h_0 \gg a$} &  \begin{tabular}{cc}
Arrhenius rates & Fermi and Bose rates \\
$\displaystyle \frac{\cosh 2a(2\mu+1) - \cosh 2a}{\cosh 2h_0} $  & \raisebox{0.15cm}{$\displaystyle  \frac{\cosh 2a(2\mu+1) - \cosh 2a}{\left \{ 1+[\alpha(1-\alpha)\Gamma \tau]^{-1}\right \}\cosh 2h_0}  $} 
\end{tabular}   \\
\end{tabular}.
\caption{Cumulant generating function $\tau \phi_\mu$ of the work (stochastic work source) during a time interval $\tau$ in the limits of fast ($\Gamma \tau \ll 1$) and slow modulation  ($\Gamma \tau \gg 1$), for small amplitudes of modulation ($a \ll 1$ ), or for large energy gaps ($h_0 \gg 1 $). We use (\ref{eq:kpmDef}) to evaluate $k^\pm$ which allows comparison with the deterministic work source (see Table~\ref{tab:limitcasesPer}). \label{tab:limitcases}}
\end{table*}
\renewcommand{\arraystretch}{1.4}
\setlength{\tabcolsep}{0.4cm}
\end{center}
%%%%%%%%%%%%%%%%%%%%%%%%

It is important to note that the only coefficient of the characteristic polynomial containing a $\mu$ dependence is the one of zero degree in $\phi$. This implies that the lowest order solutions will always contain the variable $\mu$ as expected. We also remark that if the $i$th order solution is independent of $\mu$, it must vanish since $\phi_{\mu = 0} $ is zero by definition. This is helpful to simplify the calculations leading to table \ref{tab:limitcases}. We also note the results in Table \ref{tab:limitcases} rely on the hypothesis that $\mu$ is chosen inside an interval that depends of the expansion parameter. They are valid for any types of rates except in the large field expansion ($h_0 \gg 1$) where the function $\omega(h)$ has, at large $h$, a leading role to determine the order of the various coefficient. 

%%%%%%%%%%%%%%%%%%%%%%%%%%%%%%%%%%%%%%%%%%%%%%%%%%%%%%%%%%%%%%%%%%%%%%%%%%%%%%%%%%%%%%%%%%%%%%%%%%%%%%%%%%
\subsection{Periodic work source}

\begin{figure}[h]
\begin{center}
\includegraphics[width= 7.5cm]{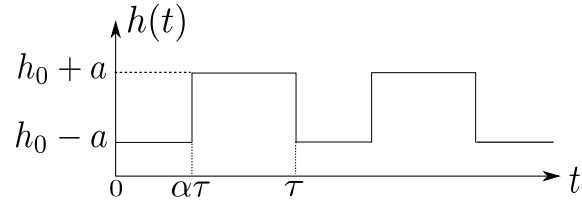}
\end{center}
\caption{Representation of the periodic work source dynamics.}
\label{fig0}
\end{figure}

We consider now the same two-level system as before but driven by the deterministic and periodic (of period $\tau$) work source depicted in Fig.~\ref{fig0}. The period fraction during which the work source is in the lower (resp. higher) state is denoted by $\alpha$ (resp. $1-\alpha$).
%%%%%%%%%%%%%%%%%
%%%%%%%%%%%%%%%%%
The large deviation function and the cumulant generating function of the work statistics are derived in \cite{Verley2013_vol88}. We briefly summarize the derivation. The work generating function $G_{\s,\mu}(t)= \langle e^{\mu W(t)} \delta_{\s,\s(t)} \rangle$ evolves according to
\beq
\partial_t G_{\s,\mu}= \sum_{\s'=\pm 1} L^{\mu}_{\s,\s'}(h(t)) G_{\s',\mu}, \label{eq:GenFuncEvo}
\eeq
where $L^{\mu}_{\s,\s'}(h) = -\s \s' \omega(h) e^{-\s' h}-\dot h \mu \s \delta_{\s,\s'} $. The asymptotic work cumulant generating function is given by the logarithm of the highest eigenvalue $\lambda_\mu$ of the following propagator
\beq
\bm{Q} =  \overrightarrow{\exp} { \int_{0}^{\tau} \bm{L}^{(\mu)}(h(t)) \D t}, \label{eq:factProp}
\eeq
where $ \overrightarrow{\exp}$ stands for the time-ordered exponential. If $g_{\s,\mu}$ denotes the eigenvector associated to the eigenvalue $\lambda_\mu$ and $g_\mu = \sum_\s g_{\s,\mu}$ the sum of its components, after $n$ periods we get
\beq
G_\mu(n \tau ) = \sum_{\s,\s'} (\bm{Q}^n)_{\s,\s'}  g_{\s',\mu} = (\lambda_\mu)^n g_{\mu},    %\asymp \exp{\left [ t\phi_\mu\right ]}.
\eeq
which leads to the asymptotic cumulant generating function
\beq
\tilde \phi_\mu = \lim_{n \rightarrow \infty}\frac{1}{n}\ln G_{\mu}(n\tau) = \ln\lambda_\mu . \label{eq:DefOfGenFunc2}
\eeq

The propagator over one period $\bm{Q}$ can be decomposed into the norm conserving evolutions over $\alpha \tau$ and $(1-\alpha) \tau$ interspersed by the propagation over the two time steps coinciding with the change of the work source state $h$. In the end, the work cumulant generating function is found to be
\beq
\fl \tilde \phi_\mu  =\ln \left ( \frac{A}{2}\cosh 2a(2\mu+1) +\frac{B}{2} \\ + \frac{1}{2} \sqrt{ \left [ A\cosh  2a(2\mu+1) +B \right ]^{2} - 4 e^{-(1-\alpha) \tau \ot^+-\alpha \tau \ot^-} } \right ) \label{eq:ScaledGenCumFunc},
\eeq
where
\bea
A & =& \frac{ ( 1 - e^{-(1-\alpha) \tau \ot^+})  (1 - e^{-\alpha \tau \ot^-})}{\cosh 2a + \cosh 2h_0}, \\ 
B & =& \frac{\left ( 1+e^{-(1-\alpha) \tau \ot^+-\alpha \tau \ot^-} \right ) \cosh 2h_0 }{\cosh 2a + \cosh 2h_0} +\frac{ \left (e^{-(1-\alpha) \tau \ot^+}+e^{-\alpha \tau \ot^-} \right)  \cosh 2a }{\cosh 2a + \cosh 2h_0}. \nonumber 
\eea
We defined $\ot^\pm = 2 \omega(h_0 \pm a) \cosh(h_0 \pm a)$ as in the stochastic work source case. The expansion of $\tilde \phi_\mu$ in the limit of fast and slow modulation and in the limit of small amplitude $a$ and large $h_0$ are given in Table \ref{tab:limitcasesPer}.
%%%%%%%%%%%%%%%%%%%%%%%
%% Tableau avec tous les cas limites CAS PERIODIQUE
\begin{center}
\renewcommand{\arraystretch}{2.6}
\setlength{\tabcolsep}{0.4cm}
	\begin{table*}[h!]
 \begin{tabular}{c|C} 
$\Gamma \tau \ll 1$ & \raisebox{0.12cm}{$ \displaystyle \frac{ \tau}{2}\sqrt{ \left [ \alpha \ot^- +(1-\alpha)\ot^+ \right ]^2 + 4 \alpha (1-\alpha) \ot^+ \ot^- \frac{\cosh 2a(2\mu+1) - \cosh 2 a }{\cosh 2a + \cosh 2 h_0 }  } $ } \\
& \raisebox{0.7cm}{$ \displaystyle
-\frac{ \tau}{2}\alpha \ot^- -\frac{ \tau}{2}(1-\alpha)\ot^+$} \\
$\Gamma  \tau \gg 1$ & \raisebox{0.12cm}{$ \displaystyle \left (e^{-(1-\alpha) \tau \ot^+} + e^{-\alpha \tau \ot^-} \right ) \frac{ \cosh 2 a - \cosh 2a(2\mu+1)}{\cosh 2h_0 + \cosh 2a(2\mu+1) }$}  \\
& \raisebox{0.7cm}{$ \displaystyle + \ln \frac{ \cosh 2 h_0 + \cosh 2a(2\mu+1)}{\cosh 2a + \cosh 2 h_0} $ } \\
$ a \ll 1$ & \raisebox{0.12cm}{$ \displaystyle \mu (1+\mu) \langle \tilde w \rangle  =  8a^2 \mu (1+\mu) \frac{(1- e^{-(1-\alpha)\tau\ot^0})(1-e^{-\alpha \tau \ot^0})}{(1-e^{-\tau \ot^0})(1+\cosh 2 h_0)} $ }  \\
\raisebox{-0.6cm}{$h_0 \gg a$} & 
			\begin{tabular}{cc}
Arrhenius rates & Fermi and Bose rates \\
$\displaystyle \frac{\cosh 2a(2\mu+1) - \cosh 2a}{\cosh 2h_0} $  & $ \displaystyle \frac{\left (1-e^{-(1-\alpha)\Gamma \tau}\right ) \left (1-e^{-\alpha \Gamma \tau}\right )}{1-e^{-\Gamma \tau}}  $  \\ 
& $ \displaystyle  \times \frac{\cosh 2a(2\mu+1) - \cosh 2a}{\cosh 2h_0}$
			\end{tabular} 
		\end{tabular}
\caption{Cumulant generating function $\tilde \phi_\mu$ of the work (periodic work source) per period in the same limits as in Table~\ref{tab:limitcases}. In the low amplitude limit, we have defined $\omega^0 = 2\omega(h_0) \cosh h_0$. \label{tab:limitcasesPer}}
	\end{table*}
\end{center}
%%%%%%%%%%%%%%%%%%%%%%%%

%%%%%%%%%%%%%%%%%%%%%%%%%%%%%%%%%%%%%%%%%%%%%%%%%%%%%%%%%%%%%%%%%%%%%%%%%%%%%%%%%%%%%%%%%%%%%%%%%%%%%%%%%%
\subsection{Thermodynamics and average behavior}

We now turn to the analysis and comparison of the work statistics generated by the stochastic and periodic work source. In both cases, we will compare the work accumulated during a time $\tau$. Since $\tilde \phi$ describes the statistics of work per period $\tau$ for the periodic work source and since $\phi$ describes the statistics of work per unit time for the stochastic work source, to compare the two, $\tilde \phi$ and $\tau \phi$ have to be considered. Furthermore, the parameters setting the time scale of the two work sources have to be related via
\beq
 k^+ =\frac{1}{(1-\alpha)\tau} \quad \mbox{and} \quad  k^- = \frac{1}{\alpha \tau },
 \label{eq:kpmDef}
\eeq
in order to spend, in average, the same amount of time at high and low value of $h$. We keep this convention throughout the paper. We also count time in unit of a period $\tau$, i.e. $t= n \tau$. 

For large $n$, in the sense of large deviation the work $w = W/n$ becomes minus the heat $q=Q/n$ because the system internal energy is bounded and thus its change between $0$ and $t$ is not extensive in time. Similarly, the entropy production per period becomes equal to the heat flow $-q$ because the system entropy change is not extensive in time. This implies that the work fluctuations fully characterize the large deviation properties of the entropy production and of the heat fluctuations. 

\begin{figure}[h]
\begin{center} 
\includegraphics[width= 7cm]{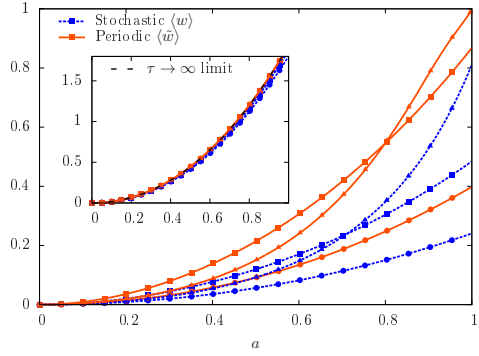}
\end{center}
\caption{Mean work value versus amplitude parameter $a$ for the periodic (orange solid lines) or for the stochastic (blue dashed lines) work sources. Symbols encode the types of rates: Arrhenius (squares), Bose (triangles) and Fermi (circle). Other parameters are $\alpha = 0.3$, $h_0=1$ and $\tau=1$ (Inset $\tau = 100 $) with rates $k^\pm$ set from (\ref{eq:kpmDef}). } 
%($k^-=3.33$ and $k^+=1.43$)
\label{fig5}
\end{figure}

For the periodic work source, the first derivative of $\tilde \phi_\mu$ at $\mu=0$ is the average work per period
\beq
\langle \tilde w \rangle = \frac{4a \sinh(2a)\left( 1-e^{-(1-\alpha)\tau\ot^+} \right) \left( 1-e^{-\alpha \tau\ot^-} \right)}{\left[ \cosh(2h_0) + \cosh(2a) \right]\left( 1-e^{-\alpha \tau\ot^- -(1-\alpha)\tau\ot^+} \right)}. \label{eq:AvgWorkPer}
\eeq
For the stochastic work source, we show in \ref{sec:AvgWorkStoch} that the work received by the two-level system during time $\tau$ is
\beq
\langle w \rangle  =  4a \tau k^+ k^-\left(\omega^-_+ \omega^+_- -\omega^+_+ \omega^-_- \right )/Z, \label{eq:AvgWorkStoch}
\eeq
with $Z$ a normalization constant in the state probability. Note that, in (\ref{eq:AvgWorkPer}) and (\ref{eq:AvgWorkStoch}), the average work is positive as required by the second law. For both types of work source, the average work vanishes for $a \rightarrow 0 $ (no work source) and for $\tau \rightarrow 0$ or $\alpha \rightarrow 0$ or $1$ (the two-level system has no time to switch state between two work source transitions). For $\Gamma \tau \rightarrow \infty $, the two-level system typically relaxes to equilibrium before the next transition in the work source happens and for both types of work source
\beq
\langle \tilde w \rangle = \langle w \rangle = \sum_{\s, \varepsilon } \frac{2a \varepsilon \s  \omega_{-\s}^\varepsilon}{\omega_+^\varepsilon + \omega_-^\varepsilon} = \frac{4a \sinh(2a)}{ \cosh(2h_0) + \cosh(2a)}. \label{eq:AvgWorkLargeT}
\eeq
$2a\varepsilon\s$ is the work given to the two-level system when the joint system state is $(\s,\varepsilon)$ and a transition $\varepsilon \ra - \varepsilon$ of the work source occurs. We notice that this average work is independent of any dynamical parameters, see Fig.~\ref{fig5}. This limit is not reversible since the driving contains discontinuities. Indeed, the average work is different from the free energy difference. In Fig.~\ref{fig5}, we present the average work for $\tau = 1$ and $\tau =100$ as a function of the amplitude of the jump $a$. We notice that the dynamics with Fermi rates always produces less work than the other rates because the Fermi rates are the smallest. Indeed, an energy exchange between the work source and the two-level system requires a change of system state between two work source transitions. Therefore the smaller Fermi rates lead to a smaller number of system transitions and thus to a smaller average work. We also see that the average work is always higher for the periodic protocol than for the stochastic one. This is due to the fact that for a Poisson process, the most likely time intervals between work source transitions are the small ones during which the system has less time to change its state and absorb work in average.
%%%%%%%%%%%%%%%%%%%%
%%%%%%%%%%%%%%%%%%%%

%%%%%%%%%%%%%%%%%%%%%%%%%%%%%%%%%%%%%%%%%%%%%%%%%%%%%%%%%%%%%%%%%%%%%%%%%%%%%%%%%%%%
\subsection{Fluctuations and statistics of work}

\begin{figure*}[h!]
\begin{center}
\includegraphics[width= 7cm]{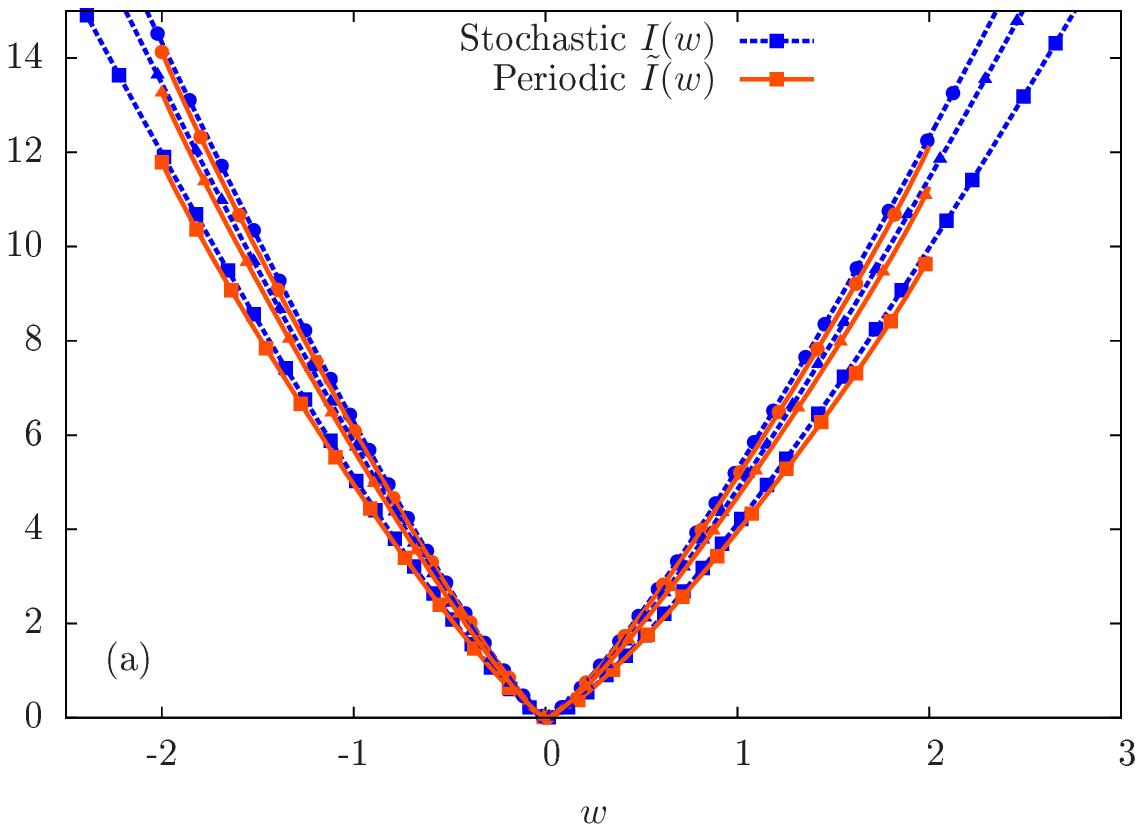} \hspace{-0.4cm}
\includegraphics[width= 7cm]{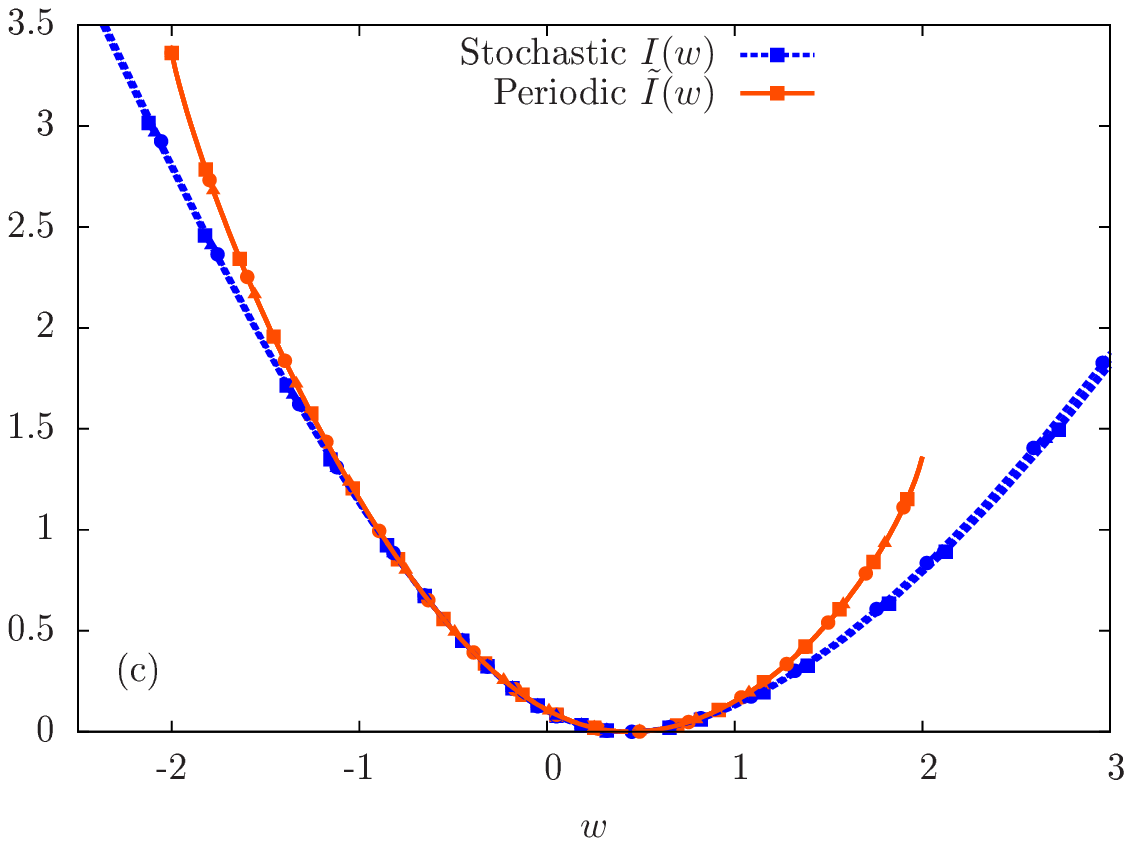}
\includegraphics[width= 7cm]{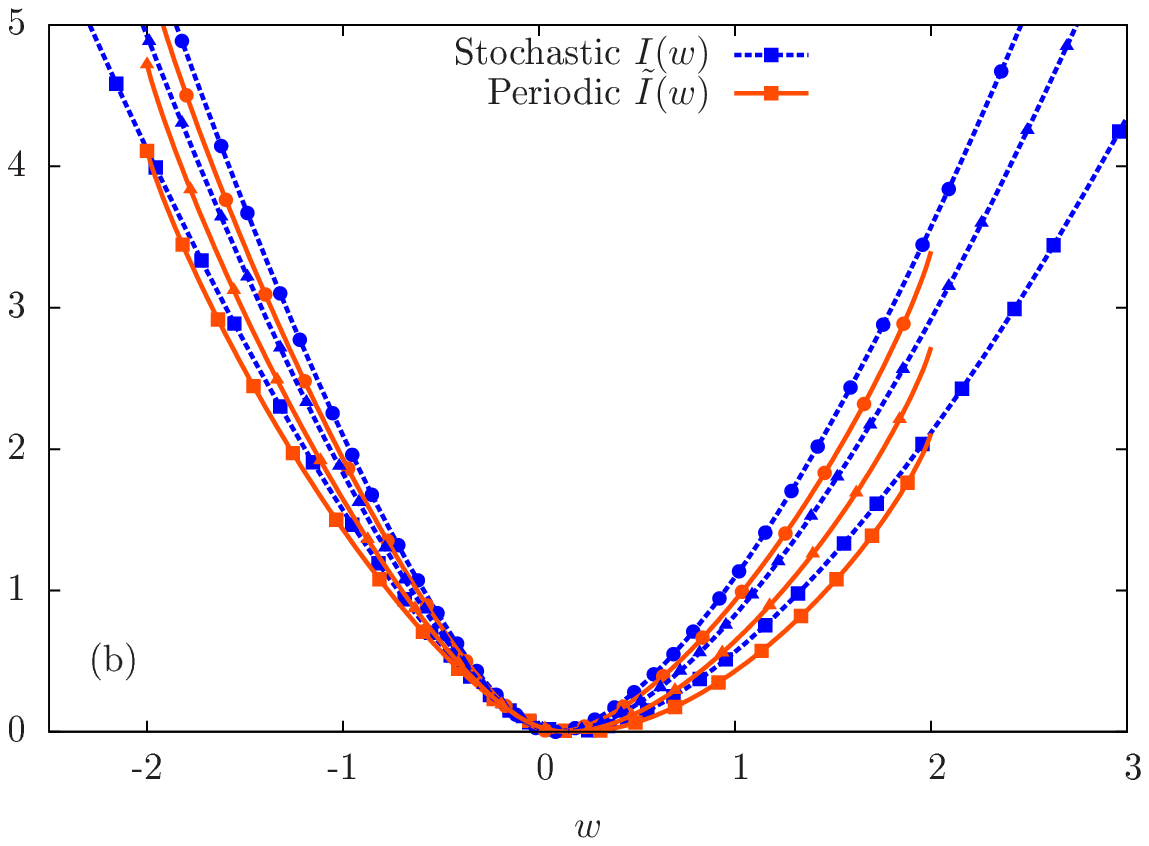} 
\caption{Large deviation functions for the work $w$ produced by the periodic (orange solid lines) and the stochastic (blue dashed lines) work source for (a) fast $\tau=0.01$, (b) intermediate $\tau = 1$, and (c) slow $\tau = 100$ switching rates compared to the system time scale. Symbols encode different types of rates: Arrhenius (squares), Bose (triangles) and Fermi (circle). Other parameters are $h_0=1$, $a=0.5$, $\alpha=0.3$. \label{fig15}} 
\end{center}
\end{figure*}

When comparing the work probability distributions corresponding to the two types of work source as in Fig.\ref{fig15}, the most striking feature is the difference in the range of the fluctuations. The stochastically driven system always has a small but finite probability to exchange a very large amount of work with the work source and as a result the support of the large deviation function $I(w) = \max_\mu \{\mu w - \tau \phi_\mu \}$ is infinite. However the periodically driven system displays a finite support because the work exchanged during a work source transition is $\pm 2a$ thus leading to three possible values for the work per period, $\pm 4a$ and $0$, and to the inequality $|w| \le 4a$. The large deviation function of work $\tilde I(w) = \max_\mu \{ \mu w - \tilde \phi_\mu \}$ is therefore infinite (vanishing work probability) outside that range. At the level of the cumulant generating function, this implies that $\tilde \phi_\mu$, the Legendre transform of $\tilde I(w)$, has an absolute slope that cannot exceed $4a$. 

%%%%%%%%%%%%%%%%%%%%
%%%%%%%%%%%%%%%%%%%%

\begin{figure}[h!]
\begin{center}
\includegraphics[width= 7.5cm]{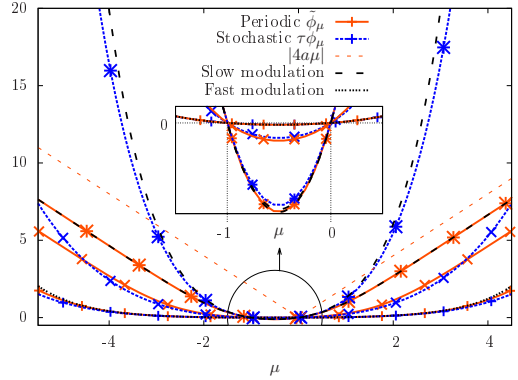}
\end{center}
\caption{Cumulant generating function of the work statistics as a function of $\mu$ for the periodic (orange solid) and stochastic (blue dashed) work source for fast $\tau=0.1$ (plus symbol), intermediate $\tau=1$ (crosses), and slow $\tau=100$ (stars) switching rates compared to the system time scale. Fermi rates are considered and $h_0 = 1$, $\alpha = 0.3$. The black dashed line corresponds to $\Gamma \tau \rightarrow \infty$ while the black dotted line to $\Gamma \tau \rightarrow 0$.}
\label{fig6}
\end{figure}

Both types of work probability distributions satisfy the fluctuation theorem. In fact any stochastic energy source made of two states is reversible and will thus qualify as work source. 
The fluctuation theorem for the periodic work source should in principle relate the work statistics for a forward periodic driving with that of the time-reversed periodic driving. However, also here a periodic driving jumping once back and forth between two states over a period is invariant under time-reversal (up to a time-shift which plays no role in the long time limit \cite{Verley2013_vol88}). 
We explicitly prove the work fluctuation theorem for the stochastic as well as for the periodic work source by showing that the work cumulant generating function satisfies the relation $\phi_\mu = \phi_{-1-\mu}$ \cite{Kurchan1998_vol31}. This follows directly from (\ref{eq:ScaledGenCumFunc}) and indirectly from (\ref{eq:cMain}) by noticing that $\mu$ only appears in the characteristic polynomial through the function $\cosh[2a(1+2\mu)]$. This function is invariant in the exchange of $\mu$ to $-1-\mu$ and this also holds true for the roots of the polynomial. This symmetry is also explicitly seen in Fig.~\ref{fig6}.
%%%%%%%%%%%%%%%%%%%%
%%%%%%%%%%%%%%%%%%%%

We now turn to the quantitative analysis of the large deviation function for work $I(w)$ obtained by numerical Legendre transform of the cumulant generating function corresponding to the two types of work sources.

Fig.~\ref{fig15}a and the tables \ref{tab:limitcases} and \ref{tab:limitcasesPer} show that both types of work sources give rise to identical work fluctuations in the limit of fast modulation ($\Gamma \tau \ll 1$). The choice of the rates nevertheless influences the shape of the work distribution and the work variance decreases as the modulation speeds up. We also identify a linear behavior in $\tau$ as $\tau \rightarrow 0$. 
In the opposite limit of slow modulation ($\Gamma \tau \gg 1$), the large deviation function becomes rate independent as seen on Fig.~\ref{fig15}c. However, the type of work source (stochastic or periodic) still influences the work fluctuations, in particular the large fluctuations away from the common minimum (same expectation value). 

In the limit of small field amplitudes (i.e. close-to-equilibrium), the cumulant generating function becomes quadratic in $\mu$: $\tau \phi_\mu = \mu(1+\mu) \langle w \rangle $ and $\tilde \phi_\mu = \mu(1+\mu) \langle \tilde w \rangle $, where the mean work values are obtained from (\ref{eq:AvgWorkPer}) and (\ref{eq:AvgWorkStoch}) by second order expansion in $a$. The average work differs for stochastic and periodic work sources and for the different rates. The same observation remains true at the level of work fluctuations. 

%%%%%%%%%%%%%%%%%%%%%%%%%%%%%%%%%%%%%%%%%%%%%%%%%%%%%%%%%%%%%%%%%%%%%%%%%%%%%%%%%%%%
\section{Conclusion}
\label{sec:Concl}

We identified the condition under which a system subjected to a stochastic driving can be seen thermodynamically as a system subjected to a work source, namely the stochastic driving protocol has to be statistically reversible or in other words its entropy production has to vanish. Under this assumption, the statistics of work and entropy production in the system are identical and a work fluctuation theorem is satisfied. 
We then compared the exact work statistics of a two-level system driven by a stochastic two-state work source with that of a periodic two-state work source, the two sources spending the same fraction of time in their upper and lower states. We found that the work fluctuations are unbounded in the former case and bounded in the latter one. For fast as well as for slow switching rate in the work source the work fluctuations are quite similar. For small switching amplitudes the system remains close to equilibrium where work fluctuations are Gaussian. Finally, for a given amplitude of the jumps in the work source, important values for the work average and variance are obtained when the time scales of the system and of the work source are comparable. \\

{\it Acknowledgements} - This work was supported by the National Research Fund, Luxembourg under Project No. FNR/A11/02 and INTER/FWO/13/09 and also benefited from the support from the ESF network "Exploring the Physics of Small Devices". 

%%%%%%%%%%%%%%%%%%%%%%%%%%%%%%%%%%%%%%%%%%%%%%%%%%%%%%%%%%%%%%%%%%%%%%%%%%%%%%%%%%
\appendix
%%%%%%%%%%%%%%%%%%%%%%%%%%%%%%%%%%%%%%%%%%%%%%%%%%%%%%%%%%%%%%%%%%%%%%%%%%%%%%%%%%
\section{Limit of the energy source}\label{LimEnSource}

In this appendix we consider a Markovian dynamics on a bipartite system with rates satisfying local detailed balance. We discuss the limiting regime utilized in section \ref{sec1} where one part of the system becomes an independent energy source for the second.

\begin{figure}[h!]
\begin{center}
\includegraphics[width= 6cm]{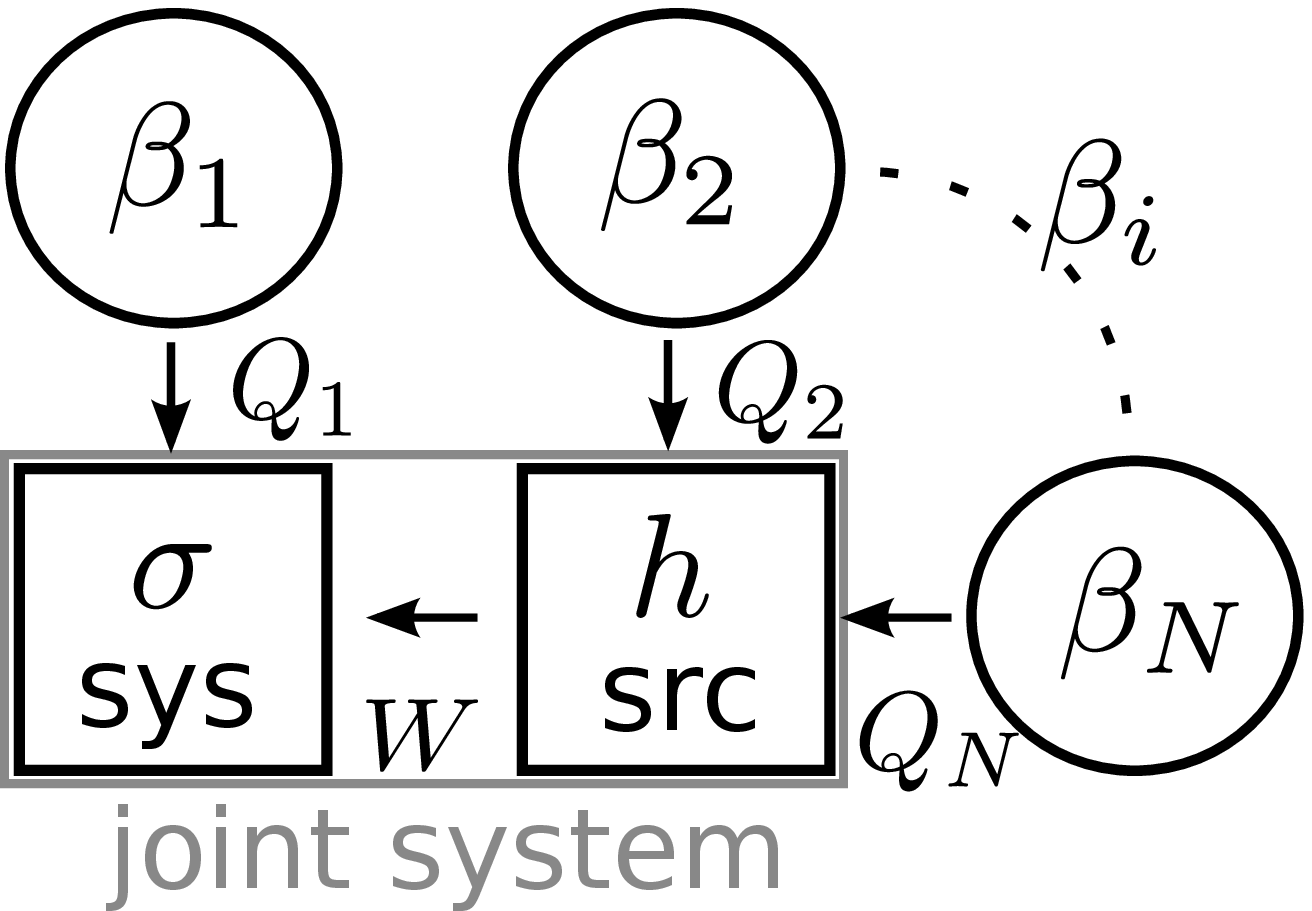}
\end{center}
\caption{The joint system $(\s,h)$ in contact with several thermal baths. In the limit where $h$ is independent of $\s$, the evolution of $h$ can be considered as a time-dependent driving for the system $\s$.}
\label{fig23}
\end{figure}

The energy of the joint system can be expressed as the sum of the bare energy of system $\s$ and $h$ plus an interaction energy, namely $E_{joint}(\s,h) = E^0_{sys}(\s)+E^0_{src}(h)+E_{int}(\s,h)$. System $\s$ is assumed in contact with a single thermal bath at temperature $\beta_1^{-1}$, while system $h$ is in contact with $N-1$ baths at temperature $\beta_\nu^{-1}$, with $\nu=2,\cdots ,N$, see Fig.\ref{fig23}.  
The rates describing transitions between states $\s$ at fixed $h$ satisfy local detailed balance
\beq
 \ln \frac{\omega^{(1)}_{\s',\s}(h)}{\omega^{(1)}_{\s,\s'}(h)} = - \beta_1 [E_{int}(\s',h)  - E_{int}(\s,h) + E^0_{sys}(\s') - E^0_{sys}(\s)], \label{eq:DBsysApp}
\eeq
as do the rates describing transitions between states $h$ at fixed $\s$
\beq
 \ln \frac{\omega^{(\nu)}_{h',h}(\s)}{\omega^{(\nu)}_{h,h'}(\s)} = -\beta_{\nu} [ E_{int}(\s,h')- E_{int}(\s,h) + E^0_{src}(h') - E^0_{src}(h)]. \label{eq:DBworksourceApp}
\eeq
We now assume that the energy scales involved during transitions in $h$ are much larger than any other energy scale involved in the $\s$ dynamics. As a result the rates $\omega^{(\nu)}_{h',h}(\s)$ can be assumed independent on $\s$ and equation (\ref{eq:DBworksourceApp}) becomes
\beq
 \ln \frac{\omega^{(\nu)}_{h',h}}{\omega^{(\nu)}_{h,h'}} = \beta_{\nu} [E^0_{src}(h') - E^0_{src}(h)] . \label{LDBautonome}
\eeq
In this limit system $h$ follows a dynamics independent from the dynamics of $\s$. However the converse is not true and the dynamics of $\s$ still depends on $h$. The energies of $\s$ read $E_{sys}(\s,h) = E_{sys}^0(\s) + E_{int}(\s,h) $, and the energy balance for $\s$ reads
\beq
\langle \dot E_{sys} \rangle = \langle \dot Q_1 \rangle + \langle \dot E^h_{int} \rangle , \label{LocEnBal}
\eeq
where the heat flow entering system $\s$ from bath $\nu=1$ is given by
\bea
\hspace{-1cm} \langle \dot Q_1 \rangle &=& \sum_{\s,\s,h} \omega^{(1)}_{\s',\s}(h) p(\s,h) \left  [ E_{int}(\s',h)  - E_{int}(\s,h) + E^0_{sys}(\s') - E^0_{sys}(\s) \right ], 
\eea
while the energy received by system $\s$ from system $h$ reads
\bea
\hspace{-1cm} \langle \dot E^h_{int} \rangle &=& \sum_{\s,\s,h,\nu} \omega^{(\nu)}_{h',h}(\s) p(\s,h) \left  [ E_{int}(\s,h') - E_{int}(\s,h) \right ].
\eea

%%%%%%%%%%%%%%%%%%%%%%%%%%%%%%%%%%%%%%%%%%%%%%%%%%%%%%%%%%%%%%%%%%%%%%%%%%%%%%%%%%
\section{Steady state probability with stochastic driving}

\label{sec:AvgWorkStoch}

We derive the average work done by the stochastic work source of Fig.~\ref{fig0bis} on our two-level system. To do so we calculate the steady probability current for the transitions leading to work exchanges. The steady state probability $p^\varepsilon_\sigma$ to find the joint system in a state $(\s,\varepsilon)$ can be obtained from the spanning tree formula \cite{Hill1989_vol} and reads
\bea
p^+_- & =& (k^-k^-\omega^+_+ + k^+k^-\omega^-_+ + \omega^+_+k^-\omega^-_+ + \omega^-_-k^-\omega^+_+)/Z, \nonumber \\
p^-_- & =& (k^+\omega^-_+k^+ + \omega^+_+k^+\omega^-_+ + \omega^+_-k^+\omega^-_+ + k^-\omega^+_+k^+)/Z,\nonumber \\
p^+_+ & =& (\omega^+_-k^-\omega^-_- + \omega^+_-k^-k^- + \omega^-_+k^-\omega^+_- + k^+\omega^-_-k^-)/Z, \nonumber \\
p^-_+ & =& (\omega^+_-k^+\omega^-_- + k^+k^+\omega^-_- + k^-\omega^+_-k^+ + \omega^+_+k^+\omega^-_-)/Z. \nonumber
\eea
$Z$ is a normalization constant such that $p^+_+ + p^-_- +p^+_- + p^-_+ = 1 $. The average work accumulated during $\tau$ is then
\beq
\langle w \rangle = 2a \tau (p^+_+ k^+ - p^-_+ k^-) + 2a\tau(p^-_- k^- - p^+_- k^+).
\eeq

%%%%%%%%%%%%%%%%%%%%%%%%%%%%%%%%%%%%%%%%%%%%%%%%%%%%%%%%%%%%%%%%%%%%%%%%%%%%%%%%%%
\section{Stochastically driven colloidal particle, the Gaussian driving case}

\label{sec:BrownPart}

We consider an overdamped colloidal particle in contact with a bath a temperature $T$ and evolving in a one dimensional harmonic trap whose position follows an Ornstein-Uhlenbeck process. Since this process plays the role of a reversible work source acting on the colloidal particle, our results of section \ref{sec1} indicate that a work fluctuation theorem should hold. However, this model has been studied experimentally in \cite{Gomez-Solano2010_vol89} and theoretically in \cite{Pal2013_vol87, Sabhapandit2012_vol85} and the Authors found that the work fluctuation theorem is only satisfied in a certain range of parameters. The reason for these violations is that work is defined in these references as the time integral of the velocity times the stochastic force and this work definition differs from the Jarzynski work by a boundary term. We now use the results of \cite{Pal2013_vol87} (trying to keep their notations) to show that, as we predicted, the Crooks work fluctuation theorem is always valid when the Jarzynski work is considered.

We denote by $x = x(t)$ the position of the particle (i.e. the system) and by $y = y(t)$ the position of the trap (i.e. the work source). The stochastic differential equations of motion are
\bea
\dot x &=& - \frac{x-y}{\tau_\gamma} + \xi, \label{eq:xMotion} \\
\dot y &=& -\frac{y}{\tau_0} + \zeta , \label{eq:yMotion}
\eea
where $\xi=\xi(t)$ and $\zeta=\zeta(t)$ are two uncorrelated Gaussian white noise averaging to zero and with correlation $\langle \xi(t) \xi(s) \rangle = 2 D \delta (t-s) $ and $\langle \zeta(t) \zeta(s) \rangle = 2 A \delta (t-s)$. Two time scales are introduced in (\ref{eq:xMotion}) and (\ref{eq:yMotion}), the relaxation time in the harmonic trap $\tau_\gamma = \gamma / k$ with $\gamma$ the friction coefficient and $k$ the stiffness of the trap, and the relaxation time of the $y$ correlation, i.e. $\langle y(t) y(s) \rangle = A \tau_0 \exp (- |t-s|/\tau_0)$.
We note that the Einstein relation $D = T / \gamma$ (which plays the role of the local detailed balance condition in equation (\ref{eq:DBsys}) when considering continuous models) is verified for the motion of $x$. The motion of the trap reaches an equilibrium (Gaussian) state which can be characterized by an effective temperature proportional to $A \tau_0$. We define $\delta = \tau_0 / \tau_\gamma$, the ratio of the two time scales in the model and $\theta = A/D$, the ratio of the diffusion coefficients.

The entropy production on the time interval $[0,t]$ contains three contributions. The first one is the variation of the system entropy
\beq
\Delta S[x|y] =-\ln \frac{p_{st}(x(t)|y(t))}{p_{st}(x(0)|y(0))} ,
\eeq
where $p_{st}(x|y)$ is the stationary probability of $x$ given $y$. 
The second one is the entropy production in the bath $-Q[x|y]/T = W[x|y]/T - \Delta E[x,y]/T $ which can be expressed as the difference between the Jarzynski work divided by $T$
\beq
W[x|y]/T = \frac{k}{2T} \int_0^t \D t' \dot y(t') \circ \frac{d}{dy}[x(t')-y(t')]^2,
\eeq
where $\circ$ denotes the Stratonovitch product, and the variation of system internal energy divided by $T$
\beq
\Delta E[x,y]/T = \frac{1}{2D \tau_\gamma} \left\{ [x(t) - y(t)]^2 - [x(0) - y(0)]^2  \right\}.
\eeq
We remark here that the Jarzynski work could be infinite for some rare events because the amplitude of change of the position $x$ is in principle infinite.
The third part in the entropy production is the work source entropy production
\beq
\Delta_{\rm i} S_{sd}[y] = - \frac{1}{2A \tau_0}  [y(t)^2 - y(0)^2] -\ln \frac{p_{st}(y(t))}{p_{st}(y(0))} = 0, \label{eq:SdEntProdColloid}
\eeq
which vanishes because the driving is a Gaussian process always relaxing to an effective equilibrium. 
Introducing the final state vector $U=(x(t),y(t))^T$, the initial state vector $U_0=(x(0),y(0))^T$, and the stationary probability distribution
\begin{equation}
p_{st}(U) = \frac{1}{2\pi \sqrt{\det \bm{H}_1} } \exp \left[ \frac{1}{2} U^T \bm{H}_1^{-1}U \right] ,
\end{equation}
with
\begin{equation}
\bm{H}_1 = \frac{D \tau_0}{\delta(1+\delta)} \left( 
\begin{array}{cc}
1+\delta+\theta \delta^2 & \theta \delta^2 \\
\theta \delta^2 & \theta \delta+\theta \delta^2
\end{array} 
\right),
\end{equation}
the entropy production becomes
\beq
\fl \Delta_{\rm i} S[x,y] = \frac{k}{T} \int_0^t \D t' y(t') \circ \dot x(t') - \frac{1}{2} U^T \left( - \bm{H}_1^{-1} + \bm{R} \right) U  
- \frac{1}{2} U_0^T \left(  \bm{H}_1^{-1} - \bm{R} \right) U_0. \label{eq:EntProdColloid}
\eeq
We defined the matrix 
\begin{equation}
\bm{R} = \frac{1}{ D \theta \tau_0} \left ( \begin{array}{cc}
\delta \theta  & 0 \\
0 & 1
\end{array} \right ).
\end{equation}
%%%%%%%%%%%%%%continue here
In Ref. \cite{Pal2013_vol87}, the generating function of the first term on the right hand side of (\ref{eq:EntProdColloid}) was obtained for a given initial and final state. The entropy production modifies this quantity by a boundary term which only depends on $U_0$ and $U$. Using the result of Ref. \cite{Pal2013_vol87}, we obtain the following expression for the generating function of the entropy production (which is equal to the dissipated work) at large time $t$:
\beq
\left \langle e^{ \mu \Delta_{\rm i} S }\right \rangle =  g_\mu e^{ t \phi_\mu }, \label{eq:GenFuncEntProdColl}
\eeq
with 
\bea
\phi_\mu &=& \frac{1+\delta }{2\tau_0} [1- \nu(\mu)], \label{eq:1} \\
\nu_\mu &=& \sqrt{1-\frac{4\theta\delta^2 \mu(1+\mu)}{(1+\delta)^2}}. \label{eq:2}
\eea
The exponential pre-factor $g_\mu$ is important if the generating function has non analyticities in the region where the saddle approximation is done to find the large deviation function.
Using again the results of reference \cite{Pal2013_vol87}, we find the following exponential pre-factor
\beq
g_\mu = \frac{4\nu_\mu}{(1+\nu_\mu)^2}.
\eeq
The work fluctuation theorem follows from (\ref{eq:1}) since $g_{-1-\mu} = g_\mu$ and $\phi_{-1-\mu} = \phi_\mu $. As for the two-level system, this theorem is satisfied because the stochastic driving is reversible, see (\ref{eq:SdEntProdColloid}). 

\vspace{1cm}
%%%%%%%%%%%%%%%%%%%%%%%%%%%%%%%%%%%%%%%%%%%%%%%%%%%%%%%%%%%%%%%%%%%%%%%%%%%%%%%%%%%%%%%%%%%%%%%%%%%%%%%%%%%%%%%%%%%%%%%%%%
\bibliographystyle{unsrt.bst}
\bibliography{/home/administrator/Documents/Gatien_pro/Programme_Java/Base_de_donnee_jabref/Ma_base_de_papier_no_url}

\end{document}